\documentclass[11pt,twoside]{article}
\usepackage{principia,endnotes}
\usepackage[charter]{mathdesign}
\usepackage[brazilian,english]{babel}
\usepackage[T1]{fontenc}

\usepackage{graphicx,float,csquotes}

\usepackage[citestyle=philosophy-classic]{biblatex}
\DeclareDelimFormat{finalnamedelim}{\addspace\&\space}
\usepackage[pdfa]{hyperref}
\newcommand{\textfoot}{}

\newcommand{\ftexthead}{\small %
\begin{tabular}{@{}p{13.5cm}}
To appear in \textsc{Principia} \scriptsize{\textbf{26}(1), 2022}\hfill{\scriptsize https://periodicos.ufsc.br/index.php/principia/index}\\

\end{tabular}}

\newcommand{\ftextfoot}{\tiny
}

\let\footnote=\endnote

\begin{document}
\thispagestyle{empty}
\frenchspacing


\begin{center}
\mbox{}

\medskip
{\lgqz\sc\bfseries Taking models seriously and being a linguistic realist}


\vspace{14pt}

\textsc{\lgtrz Raoni Wohnrath Arroyo}

\vspace{4pt}

{\scriptsize\ita{Centre for Logic, Epistemology and the History of Science, University of Campinas. Support: grant \#2021/11381-1, São Paulo Research Foundation (FAPESP). Research Group in Logic and Foundations of Science (CNPq). International Network on Foundations of Quantum Mechanics and Quantum Information,} \textsc{Brazil}\\
\texttt{rwarroyo@unicamp.br}

}


\vspace*{4pt}

\textsc{\lgtrz Gilson Olegario da Silva}

\vspace{4pt}

{\scriptsize\ita{Department of Philosophy, Federal University of Juiz de Fora,} \textsc{Brazil}\\
\texttt{gilsonolegario@gmail.com}
}

\end{center}

\selectlanguage{english}
\noindent\rule[1mm]{135mm}{.3mm}

{\small

\noindent\textbf{Abstract.}\quad
Carnap's conception of linguistic frameworks is widespread; however, it is not entirely clear nor consensual to pinpoint what is the influence in his stance within the traditional realist/anti-realist debate. In this paper, we place Carnap as a proponent of a scientific realist stance, by presenting what he called ``linguistic realism''. Some possible criticisms are considered, and a case study is offered with wave function realism, a popular position in the philosophy of quantum mechanics.
}

\medskip

{\small 


}

\noindent\rule[1mm]{135mm}{.3mm}


\noindent{\scriptsize\textsc{received:} 15/10/2021 \hspace*{1cm} \textsc{revised:} 05/04/2022 \hspace*{1cm} \textsc{accepted:} 05/04/2022 }


\section{Introduction}\label{sec:1}
Scientific realism is a philosophical position roughly characterised by the claim that scientific theories describe the world. What this amounts to is up for grabs, since pretty much every author working on scientific (anti-)realism has its own version of it \parencite[cf.][sec.~1.1]{chakravartty2017sep}. \textcite[p.~179]{putnam1975}, attributing to Boyd, states that the minimum criteria for a view to be called ``[scientific] realist'' is to state that the terms of the scientific language refer, and the scientific laws are true---motivating the former by means of the latter. Fairly enough, scientific realism has been described as the conjunction of epistemic, ontological, and syntactic components (belief, existence, and truth, respectively);\footnote{Cf. \textcite{chakravartty2017sep} and \textcite[xvii]{psillos1999}.} on the other hand, anti-realism has been presented as a denial (of a set) of the three aforementioned realist aspects. So, for instance, a textbook scientific realist would state that if non-relativistic quantum mechanics says that there are electrons, then we should \textit{believe} in its \textit{existence} because the term \textit{truly corresponds} to such an entity in the world. The anti-realist stance has, at first, a relatively easier job, consisting in maintaining that we should \textit{not believe, but accept} the existence of such an entity because it is only the \textit{convention} that is used for \textit{empirical adequacy} of mature scientific theories, such as non-relativistic quantum mechanics. 

Of course, this is a polarised view of the debate, concerning what \textcite[chap.~3]{putnam1981} called the ``God's Eye View'' on the side of scientific realism, whereas the latter concerns an instrumentalist account within the anti-realist stance. Things are not this black and white, however. There is a whole spectrum in the grey area between these two, and the debate is not even close to being settled \parencite[cf.][]{mizrahi2020}. Our working hypothesis in this article is that this happens because both scientific realism and anti-realism yield a ``yes'' or ``no'' kind of answer. This is where Carnap's works kick in for the rescue. Carnap's Principle of Tolerance and the conception of linguistic frameworks may provide a useful \textit{via tertia} on the current debate. The problem is that such contribution is often unappreciated because of several misunderstandings concerning Carnap's positions, both in the philosophy of science and ontology.

It is precisely these misunderstandings that we intend to [or contribute to] dispel in this article, which is structured as follows. Section \ref{sec:2} deals with the ``received view'' of Carnap's works, and his famous associations with ``neutralist'' and ``conventionalist'' theses concerning theory choice in the scientific frameworks/models. Section \ref{sec:3} brings textual reference of Carnap's own works to counter that received view. Building upon what was discussed previously, section \ref{sec:4} advances a realist reading of Carnap's works. In particular, we discuss some open problems in the philosophy of quantum mechanics that instantiate the Carnapian metaontology. As our working example, we choose wave function realism. Section \ref{sec:5} wraps it all up with some considerations on the traditional Carnap--Quine debate.

\section{The received view(s) of Carnap's view}\label{sec:2}

Let us begin with an example. Consider non-relativistic quantum mechanics, a physical theory that describes phenomena at the nanoscopic level. In such a quantum-mechanical description, the use of linear differential equations is fairly undisputed \parencite[cf.][]{arroyo-olegario2021}, and such equations (usually the Schrödinger Equation) often describe the evolution of the wave function \parencite[cf.][]{ney2021} in several experiments, such as the Stern--Gerlach and related spin experiments \parencite[cf.][]{maudlin2019}. By taking quantum mechanics at face value, we seem to be in a position to make the scientific-realist checklist: 
\begin{description}
    \item[Ontology:] Are there wave functions? Yes.
    \item[Epistemology:] Should we believe in its existence? Yes.
    \item[Semantics:] Is the wave-function description true? Yes.
\end{description}

But what makes the scientific realist \textit{entitled} to respond affirmatively to such questions? After all, as anti-realists rightly point out, many quantum theories do not refer to wave functions whatsoever \parencite[cf.][]{bokulich2020, allori2020}. And so the (seemingly endless) debate concerning underdetermination begins \parencite[cf.][]{arenhartarroyo2021meta}. Enter the Carnapian frameworks.

In a textbook presentation, the story goes like this \parencite[cf.][chap.~4]{ney2014}:
a linguistic framework contains linguistic expressions and a set of rules which evaluate whether a linguistic expression is true or false. Take, for example, the following expression.

\begin{quote}
    The wave function $|\psi(\textrm{x})\rangle$ of a particle in a box has the form of: $$|\psi(\textrm{x})\rangle=\sqrt{\frac{2}{a}}\textrm{sin}\left(\frac{n\pi}{a}\textrm{x}\right)$$
\end{quote}

It is a linguistic expression within the linguistic framework of quantum theories that admits wave functions. And it is a true linguistic expression inasmuch as there is \textit{some} axiomatic system  \parencite[cf.][]{arroyo-olegario2021} that guarantees its veracity. These are the internal issues, namely questions that are internal to each specific linguistic framework. Issues lying outside each specific framework would be meaningless from the point of view of such a framework. For example, take an existential/ontological question typical of realism (\textit{viz.} whether the theoretical terms refer to or not), such as ``are there \textit{really} wave functions?'' This kind of question can be internal or external to the framework. If the existential question is internal, the answer is almost trivial: wave functions do exist inside the linguistic framework of the quantum theories that admit wave functions. They are part of the linguistic expressions of such framework, \textit{i.e.}, of the expressions postulated by the language of that framework. But this is not what existential questions are. Rather, when one makes an ontological question about, \textit{e.g.}, wave functions, one wants to know whether wave functions exist \textit{in the world}, independently of whatever framework, just like tables and chairs. But then, one is asking an external question, \textit{viz.} a question that is external to such linguistic framework, which is meaningless---\textit{i.e.}, one is asking a question that goes beyond the framework in which the entity in question is postulated. 

The same goes for all existential/ontological questions. In this sense, ontology must be meaningless---and so is the quest for scientific realism (as it asks ontological/external questions). Before we proceed, a remark is in order. Normally, the term used by Carnap (and by part of the literature discussing Carnap's work) is ``metaphysics''. We are employing a distinction between ``metaphysics'' and ``ontology'' by their subject matter as offered in \textit{e.g.} \textcite{hofweber2016}: ontology deals with existence questions, and metaphysics deals with questions of nature \parencite[for more details and discussion, cf.][]{arenhartarroyo2021manu}. For example, here's Hofweber:

\begin{quote}
    In metaphysics we want to find out what reality is like in a general way. One part of this will be to find out what the things or the stuff are that are part of reality. Another part of metaphysics will be to find out what these things, or this stuff, are like in general ways. Ontology, on this quite standard approach to metaphysics, is the first part of this project, i.e. it is the part of metaphysics that tries to find out what things make up reality. Other parts of metaphysics build on ontology and go beyond it, but ontology is central to it \textelp{}. Ontology is generally carried out by asking questions about what there is or what exists. \parencite[p.~13]{hofweber2016}.
\end{quote}

Still, there is evidence that Carnap uses the term ``ontology'' with caution; in \textcite[933]{CARNAP1963-orig}, for example, he questions Beth's use of the expression ``ontological commitment'', and also had the trouble including a footnote to \textcite{CARNAP1950} to express his dissatisfaction with Quine's use of ``ontology'' \parencite[cf. also][p.~175 for further references and discussion]{olegariotese2020}. \textcite[p.~316]{davidson1963method} also points to this same caution regarding the traditional notion of the term: ``Carnap agrees with Quine's \textit{dictum} on this point (although he balks at the word `ontology')'';\footnote{We'll come back to this issue in section \ref{sec:5}.} To be on the safe side, \textcite{parrini1994carnap} suggested the term ``ontic'' for this type of question in Carnap:

\begin{quote}
    \textelp{} we can conjecture that for Carnap it is possible to reject any ``ontological commitment,'' in the metaphysical sense of this expression, without being compelled by this to reject any ontological commitment in the empirical sense of this expression (a commitment that we could call \textit{ontic}). \parencite[p.~260, original emphasis]{parrini1994carnap}.
\end{quote}

Now, it seems safe to say that the Carnapian standpoints (both in the philosophy of science and ontology) have been widely (mis)understood, and this can be easily seen with the following showcase of cross-attributions. 

\begin{itemize}
    \item \textcite[p.~45]{psillos1999} initially attributes to Carnap the defence of a special kind of structural realism, and later places him as a halfway between scientific realism and anti-realism \parencite[cf.][]{psillos2011}.
    \item \textcite[p.~10]{gentile2005}, following Quine, qualifies Carnap as a ``Platonic realist'' due to supposed ontological commitments made with reference to abstract mathematical entities in his proposal for the reinterpretation of scientific theories.
    \item \textcite[p.~69]{maudlin2007} places Carnap within the anti-realist camp, arguing that empiricist to the point of stating that ``\textelp{} Carnap is just Hume warmed over and updated''.
    \item \textcite[p.~78]{chalmers2009} calls him the first ``ontological anti-realist'', since Carnap responded negatively to the metaontological question, \textit{viz.} whether there is a fact of the matter about what exists in reality (what is the \textit{correct} framework), and \textcite[p.~130]{eklund2009} calls him an ``ontological pluralist'' for the same reason.
    \item \textcite[pp.~98, 105]{friedman2012a} suggests that Carnap is successful in establishing a neutral position with regards to the debate concerning realism and anti-realism in the philosophy of science. This is also the position of \textcite[p.~ix]{falguera-martinezvidal2020pref}, who classify Carnap as holding a ``neutral'' position between realism and anti-realism in ontology (\textit{e.g.} concerning abstract objects).
    \item \textcite[p.~68]{demopoulos2013} qualifies him as favouring an anti-realist position, and \textcite[sec.~4.1]{chakravartty2017sep} goes further arguing that he is an instrumentalist because statements about unobservable entities and processes are devoid of truth-value, therefore meaningless.
    \item \textcite[p.~122]{thomasson2016} classifies him as defending a ``\textelp{} form of ontological deflationism'', where ontological disputes are meaningless or merely verbal.
    \item \textcite[p.~343]{bueno2016} considers that Carnap's \textit{Aufbau} is a ``blend of logicism \textit{and} conventionalism'' in order to ``avoid ontological commitment to mathematical entities''---which can be generalised to any theoretical terms in science, such as ``wave functions''.
    \item \textcite[p.~2]{jaksland2020} considers Carnap to be a ``metaphysical deflationist'' because Carnap allegedly ``challenges the objectivity or framework-independence of metaphysics''.
\end{itemize}

The list could go on. Of course, be it towards ontology or the scientific endeavour, realism and anti-realism are opposite accounts. Therefore one cannot consistently endorse \textit{both} at the same time. There are two ways to go from here. On a first route, one can bite the bullet and argue that Carnap's philosophy indeed admits some level of inconsistency by maintaining anti-realism and realism at the same time towards the existence of entities posited by scientific theories. Alternatively, on a second route, one can withhold such inconsistency by pointing out that Carnap's own received views are misleading. In the next session, we follow the latter route.

\section{Language systems}\label{sec:3}

One can safely state that Carnap's position on the realist \textit{versus} instrumentalist dispute had reasonable stability of treatment, at least concerning his general position, which can be traced back to \textcite{CARNAP1928btrans} in the chapter ``Application to the realism controversy'' and \textcite{CARNAP1928atrans} in the chapter ``The metaphysical problem of reality''. The first component includes a negative part: the logical positivist reaction against ontology. For Carnap, both the realist and the instrumentalist stance, as traditionally defended, include ontological assumptions. However, contrary to what we identified as the ``received view of Carnap's view'', he does not adopt an attitude of complete rejection of the ``classical intuitions'' that motivates the two theses because of the negative diagnosis. Thus, instead of a simple refusal of ontological inferences, Carnap's option is to reinterpret them as practical motivations. At this point, we are in agreement with the analysis of \textcite[p.~39]{kraut2021rudolf}, who states that:

\begin{quote}
    Not all metaphysics is with dismissiveness: statements about the existence of systems of entities are essential to semantic theory---a vital enterprise. Carnap's `metalinguistic' pragmatism aims to legitimize ontology---at least those portions of ontology required by semantic inquiries. The meaning of ontological claims is to be understood in expressive terms: not as expressions of commitment to a way of life, but to the value and/or pragmatic advisability of deploying specific linguistic/conceptual frameworks. \parencite[p.~39]{kraut2021rudolf}.
\end{quote}

Another relevant component of this reaction to the traditional ways of presenting realism and anti-realism as ontological theses is Carnap's standard recipe presented in \textcite{CARNAP1950}, that is, his approach to comparative analysis of frameworks. That's the positive aspect of his overall stance. The notion of a linguistic framework is central to Carnap's general perspective of systematic reflection on the structure of languages and is closely related to the model of treatment of traditional philosophical problems adopted by him---a centrality that places him as one of the initiators of the ``linguistic turn'' in the philosophy \parencite[p.~251]{Neuber2014}.\footnote{For a comprehensive analysis of the notion of linguistic framework within the Carnapian perspective see \textcite{torfehnezhad2017carnap}.} In \textcite[p.~68]{CARNAP1963a} he sets out his general attitude as follows:

\begin{quote}
    Our task is one of planning forms of language. Planning means to envisage the general structure of a system and to make, at different points in the system, a choice among various possibilities, theoretically an infinity of possibilities, in such a way that the various features fit together and the resulting total language system fulfills certain given desiderata. \parencite[p.~68]{CARNAP1963a}.
\end{quote}

Although used initially in \textcite{CARNAP1950} and rarely in other works \parencite[cf.][p.~5]{torfehnezhad2017carnap}, it is reasonably uncontroversial that equivalent notions of ``linguistic frameworks'' can be found in other texts by other terms \textit{e.g.} ``language system''. Of particular importance is the presentation contained in \textcite{CARNAP1939} of the two possible methods of building a language system.

To represent the language of science---particularly physics---Carnap presents two proposals, or methods of construction.\footnote{What follows is a brief presentation of a fuller argument available in \textcite{olegariotese2020}.} There seems to be a progressive preference for the second form over the first. However, it is possible to both identify Carnap as favouring one form or another, depending on the particular historical focus assumed for classification. Those who focus on Carnap's first proposals will recognise him as closer to traditional empiricism/anti-realism. Those who focus on his later work will find a more distant version of traditional empiricism, in a liberalised form. The difference between the two alternative versions can be clarified by analogy to Einstein's maxim contained in ``Geometry and Experience'' \parencite{Einstein1921}---which Carnap appropriates in a particular way, extending to the structuring of his alternative methods: ``as far as the propositions of mathematics refer tocons reality, they are not certain; and as far as they are certain, they do not refer to reality.'' \textcite{CARNAP1966} repeats this analogy in another context:

\begin{quote}
    Einstein spoke of ``mathematics'', but he meant geometry in the two ways that it can be understood. ``So far as theorems of mathematics are about reality'', he said, ``they are not certain.'' In Kantian terminology, this means that they are only synthetic, they are not \textit{a priori}. ``And so far as they are certain'', he continued, ``they are not about reality.'' In Kantian terminology, only they are \textit{a priori}, they are not synthetic. \parencite[p.~183, original emphasis]{CARNAP1966}.
\end{quote}

The aforementioned ``two methods'' are presented in \textcite{CARNAP1939} with the aid of the diagram reproduced here as Figure \ref{fig:diagram}:

\begin{figure}[H]
  \centering
  \includegraphics[scale=0.2]{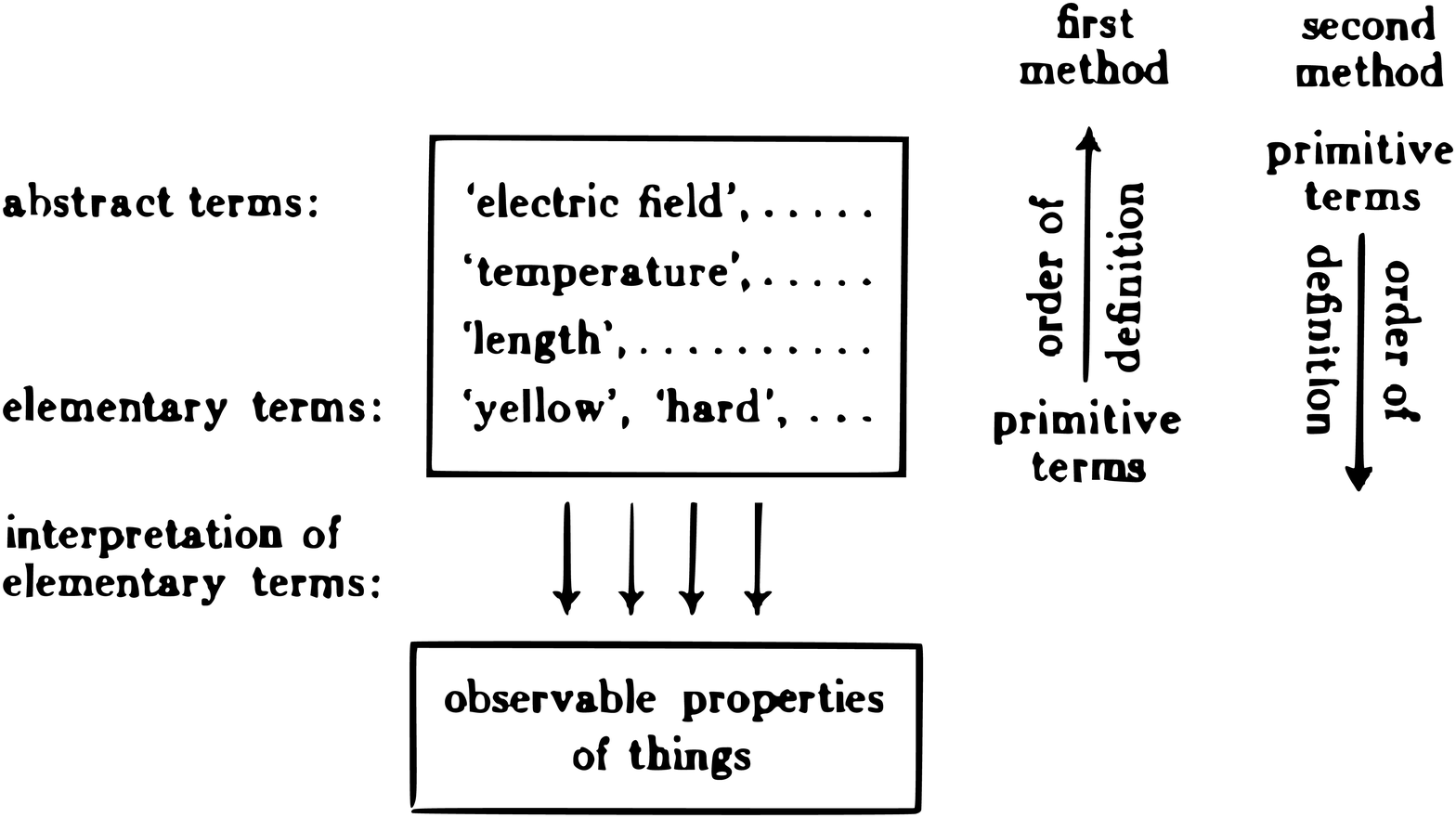}
  \caption{Diagram of the linguistic system \parencite{CARNAP1939}.}
  \label{fig:diagram}
\end{figure}

Briefly put, the first alternative for the construction of a language system construction is called the ``bottom-up method'', consisting in adopting the most elementary terms (yellow, hard, \dots) as primitive terms, and building the additional abstract terms (temperature, electric field, \dots) based on the elementary ones. But \textcite[p.~35]{CARNAP1938b} acknowledges that ``[t]his, however, is not an assertion but a proposal; a psychological basis can certainly also be chosen (and has been used in a former book of mine)''. This first method, says Carnap, is useful for teaching physics to a layperson:

\begin{quote}
    \textelp{} suppose we have in mind the following purpose for our syntactical and semantical description of the system of physics: the description of the system shall teach a layman to understand it, i.e., to enable him to apply it to his observations in order to arrive at explanations and predictions. \parencite[p.~62]{CARNAP1939}.
\end{quote}

That is, a layperson can ``understand'' the most basic physical relationships, and explain and predict phenomena endowed only with normal perceptual abilities and a minimal understanding of the language in which observational properties are described; and, progressively, can build more complex physical relations. In order to pursue this \textit{desideratum} of basic comprehensibility for the layperson, the first method will avoid the attribution of semantic rules to the more abstract terms of the scale. The system then starts from the assignment of semantic rules only those most elementary terms---which are assumed to be primitive---and connects them to the observational properties of things; then, step by step, the linguistic system is built up to the point of making it possible to understand even the most abstract terms. This first method is similar to the implementation of the ideal of the sensationalist form of science promoted by Goethe in the classical polemic against Newton, as well as that promoted by some classical positivists \parencite[cf.][p.~64]{CARNAP1939}.

Although advantageous with respect to a clear exposition of the empirical rationale and ease of understanding, the first method is not effective in promoting a potent physical system---or, as stated by \textcite[p.~64]{CARNAP1939}, ``\textelp{} it turns out -- this is an empirical fact, not a logical necessity -- that it is not possible to arrive in this way at a powerful and efficacious system of laws''. Thus, as much as historically this more simplified version (\textit{e.g.} concentrating on formulations in less complex terms) has been adopted, the discovery of counter-examples and exceptions confined the validity of laws to increasingly restricted domains. This naturally favoured the tendency of the scientific community to organise itself according to the second method \parencite[cf.][p.~64]{CARNAP1939}. 

The second method, named the ``top-down method'', expresses different \textit{desiderata} from the first, focusing on the explanatory scope of the phenomena. For this, it assumes few terms with a high degree of abstraction as primitive and a few corresponding laws with great generality, in which the elementary terms are deductively obtained by them. The semantic rules responsible for grounding the system in the observations only have an indirect relationship with primitive terms, produced by a chain of definitions that goes from abstract terms to elementary terms. With the constructed calculus ``floating in the air'', it is ``constructed downwards'', and through this chain of definitions it is finally linked down to the elementary terms: ``The laws, whether general or special, are not directly interpreted, but only the singular sentences.'' \parencite[p.~65]{CARNAP1939}. These two methods/models, of course, are simplifications and are not purely and straightforwardly applied. For example, it often happens that definitions that are restricted to abstract terms for some abstract and general laws are unknown, and in that case, they need to be assumed to be primitive. However, the most advanced fields of science apply the second method with relative success \parencite[cf.][p.~34]{CARNAP1938b}.

Note, however, that the restriction placed by Carnap on the first method is an \textit{empirical fact}---that is, that it does not produce an efficient system for the most recent theories of physics. Both methods are constructions of linguistic systems for theories, and therefore, for the system to be effective, there is no possibility of complete arbitrariness since we are, to fulfil his purpose, empirically constrained. But also note that the same is true for the second method. There is a negotiation of a kind of regime of weighing losses and gains. Although the first method is advantageous to some extent, as it clearly exposes the empirical basis and facilitates its understanding, in the long run, its efficiency reaches a limit. There is, however, no ``logical impediment'' of the process being developed by the first method, just that, if it is done so, the complexity and quantity of the laws increases (by the number of restrictions and exceptions for each counterexample).\footnote{The same idea of weighing gains and losses and the absence of ``logical impediments'' is explored later in \textcite{CARNAP1950}.}

To understand how a scientific theory works, one must only provide a partial interpretation of the abstract (theoretical) that are posited. However, when demanding an understanding of a physical theory, this understanding must mean the ability to describe or to predict new facts. This, in turn, can be provided by either the first or the second method. But if an ``intuitive understanding''---or a direct translation of an expression in terms of observational properties---is required, for Carnap, this is not possible and much less necessary. If the classification established by the distinction between theoretical terms and observational terms is linguistic, then it is certainly a conventional one. And, as such, it is more adequate or less adequate---but not more or less \textit{right}. If the aforementioned adequateness depends on a reality that is somehow essentially independent, then the refusal of this possibility (\textit{e.g.}, because it does not account for a ``support in reality'') demonstrates that such a requirement expresses a realist demand, which awaits that an independent reality presents the separation between the theoretical and the observational.

However, the idea that observational terms are understood as essential or independent properties of any linguistic framework is something that Carnap expressly rejects. The distinction is bound to be criticised for its inadequacy of this ideal, but not for the possibility of determination---unless if one claims that the determination of the language perfectly represents a substantively realistic world. If such claim is abandoned, the distinction regains its value as an artificial methodological distinction for the analysis of the language of science which is, in turn, expressed by a stipulation; and, as such, it may or may not be \textit{used}.

The hesitation to assert ``reality'' for theoretical language, \textit{i.e.}, the caveat that the situation is more complicated is \textit{precisely} due to reservations about traditional metaphysical formulations. Thus, the acceptance of the theoretical postulates of a linguistic framework is accompanied by the joint acceptance of a given interpretation---which goes back to the distinction between internal and external issues. If it understood as intended by \textcite{CARNAP1950}, the issue of the reality of theoretical terms is virtually equivalent to accepting a proposal for a form of language, and this encompasses theoretical terms that have certain definitions and relations with observation instituted by the framework's correspondence rules.

Questions concerning the independent existence of any linguistic framework have no cognitive meaning but can be turned into meaningful questions and re-established with scientific meaning if understood as equivalent to the acceptance of a language. The notion of (partial) interpretation, and therefore the recognition of the role of correspondence rules in deriving observational consequences for a theory, is part of the condition for understanding the notion of the existence---\textit{viz. internal existence}---of theoretical terms.

\section{Lessons learned: linguistic realism}\label{sec:4}

With that said, the first lesson learned from the Carnapian metaontology is, \textit{contra} the deflationist account, that \textit{ontology matters}. Ontology plays a role in semantic discussions when we want to investigate what exists \textit{inside} or \textit{modulo} each linguistic framework.\footnote{As this is a pressing issue, it is important to emphasise---as we already did---that Carnap has reservations about the use of the word ``ontology''. For instance, in \textcite[p.~43]{CARNAP1947} he states: ``I should prefer not to use the word `ontology' for the recognition of entities by the admission of variables. This use seems to me to be at least misleading; it might be understood as implying that the decision to use certain kinds of variables must be based on ontological, metaphysical convictions''; nevertheless, he concedes in the previous sentence that such a dispute could be ``of a merely terminological nature'' \parencite[cf. also][p.~50]{olegariotese2020} with regards to the preferred way to employ the terms ``ontology'' and ``metaphysics''.} In this sense, just because it doesn't go beyond the framework of each theory, that doesn't mean that ontology doesn't matter at all. In fact, that's what the use of the word ``deflationary'' suggests. A quick glance at several dictionaries teaches us the following. The verb ``deflate'' means: ``to show that (something) is not important or true'' \parencite{bd:deflate}; ``to reduce in size, importance, or effectiveness'' \parencite{mw:deflate}; ``[t]o reduce the size or importance of (a thing). Of a person's reputation, character, etc.: to depreciate, to `debunk''' \parencite[386]{oed1989}; in a figurative sense, ``[s]omeone or something that is deflated suddenly feels or is considered less important'' \parencite{ced:deflate}. If this is what people usually have in mind when describing Carnap's approach to ontology, then we seem to lost something important in the way. 

This brings us even closer to \textcite[p.~33]{kraut2021rudolf}: ``Carnap's goal is not to let the air out of ontology and minimize it, but rather to portray it as legitimate in the face of empiricist misgivings. He wishes to earn ontology the right to go on''. That is, not \textit{every} ontological endeavour should earn these rights to go on, but precisely---and exclusively---\textit{those connected with the scientific endeavour}. This is what brings the Carnapian accounts in metaontology with the debates concerning scientific realism. Here's Carnap:

\begin{quote}
    The realistic language, which the empirical sciences generally use, and the constructional language have actually the same meaning: they are both neutral as far as the decision of the metaphysical problem of reality between realism and idealism is concerned. It must be admitted that, in practice, \textit{linguistic realism,\footnote{It is worth emphasising that the term ``linguistic realism'' is a term coined by Carnap himself \parencite[86]{CARNAP1928atrans}.} which is very useful in the empirical sciences, is frequently extended to a metaphysical realism; but this is a transgression of the boundary of science} \textelp{}. \parencite[p.~86, emphasis added]{CARNAP1928atrans}.
\end{quote}

We take it that this point is crucial for the understanding of an ``irenic'', ``diplomatic'' or ``conciliatory'' (but definitely not ``neutralistic'') position of Carnap's standpoint on ontology, which can be seen as a proposal for the dissolution of the clash between scientific realism and anti-realism.\footnote{The history of the problem involves a journey through a different treatment of the interpretation of theoretical terms and special use of Ramsey sentences for the definition of analyticity in theoretical languages \parencite[cf.][]{olegariotese2020}.} If Carnap belittles the influence that ontology can have on the presentation of theories---and the use of expressions such as ``pseudo-problems'' \parencite{CARNAP1928btrans} and ``overcoming'' \parencite{CARNAP1931} does not allow us to conclude otherwise---this same depreciation is significantly unobtrusive in his later texts: the concern is with the potential impact that transgression, that is, with the influence that mental representations that accompany scientific statements, can play in the inter-subjective accountability of scientific discourse.

Linked to the qualification of (meta)ontological deflationism, there is the characterisation of the Carnapian stance as that of \textit{neutrality} in relation to the debate between realism and anti-realism, as defended by \textcite[chap.~3]{psillos1999}---also characterised as a stance which stands midway between realism and anti-realism, allegedly motivated by \textcite{CARNAP1966}. Below, we bring textual evidence for Carnap's standpoint on the matters concerning scientific realism. As it happens, his posture would be better described as prudence, or even as a systematic suspension of judgement concerning such a debate. Let us begin with how Carnap himself frames the matter.

\begin{quote}
    It is true that physicists find it vastly more convenient to talk in the shorthand language that includes theoretical terms, such as ``proton'', ``electron'', and ``neutron''. But if they are asked whether electrons ``really'' exist, they may respond in different ways. \parencite[p.~254, original emphasis]{CARNAP1966}.\end{quote}

A first way to respond, says Carnap, would be that of anti-realism \textit{qua} instrumentalism, \textit{viz.} the view according to which:

\begin{quote}
    \textelp{} theories are not about ``reality''. They are simply language tools for organizing the observational phenomena of experience into some sort of pattern that will function efficiently in predicting new observables. The theoretical terms are convenient symbols. The postulates containing them are adopted because they are useful, not because they are ``true''. They have no surplus meaning beyond the way in which they function in the system. It is meaningless to talk about the ``real'' electron or the ``real'' electromagnetic field. \parencite[p.~255]{CARNAP1966} 
\end{quote}

As opposed to anti-realism towards theoretical terms and (unobservable) entities postulated by scientific theories---e.g. ``electrons''---there is the realist account, which takes these entities to be ``really'' existent. It is worth noticing that such choice is presented by Carnap as being based on psychological grounds:

\begin{quote}
    Advocates of this approach find it both convenient and psychologically comforting to think of electrons, magnetic fields, and gravitational waves as actual entities about which science steadily learning more. \textelp{} Proponents of the descriptive view remind us that unobservable entities have a habit of passing over into the observable realm as more powerful instruments of observation are developed. \textelp{} \parencite[p.~254, original emphasis]{CARNAP1966}. \end{quote}

After presenting both, Carnap proceeds with an assessment of anti-realism and realism. To do so, he acknowledges that these are two opposite views. But their difference is linguistic. To some readers this might imply that such disagreement, by being merely linguistic, is \textit{verbal} as opposed to \textit{substantial}; hence, could be \textit{deflated} in the above-mentioned sense. But as soon as one reminds that Carnap self-proclaimedly advocates \textit{linguistic realism}, and that language means ontology, then such a reading begins to sound uncharitable to say the least.

\begin{quote}
        To say that a theory is a reliable instrument---that is, that the predictions of observable events that it yields will be confirmed---is essentially the same as saying that the theory is true and that the theoretical, unobservable entities it speaks about exist. Thus, there is no incompatibility between the thesis of the instrumentalist and that of the realist. At least, there is no incompatibility so long as the former avoids such negative assertions as, ``\textelp{} but the theory does not consist of sentences which are either true or false, and the atoms, electrons, and the like do not really exist''. \parencite[p.~256]{CARNAP1966}.\footnote{See \textcite[\S 3.4]{olegariotese2020}.}
\end{quote}

The problem at hand, which lies in the heart of the disagreement between anti-realists and realists, is that the ``reality'' of unobservable entities/theoretical terms lies \textit{outside} of the frameworks. They're external questions. But when taking into account ontological matters \textit{inside} the linguistic frameworks, anti-realists and realists stand together rather than in opposition. They're on the same boat, so to speak. The textual evidence brought above surely paths the way toward a diplomatic approach to realism and anti-realism in the ontology of science. But there is no ``neutral'' position here, as their difference is acknowledged. Such a difference can be diminished within linguistic frameworks, hence Carnap's linguistic realism.

So if ontological questions, \textit{viz.}, existence questions, are not deflated, the second lesson learned concerns the ontological aspect of scientific realism: are there entities posited by scientific theories? In the light of what was presented, let us revisit wave function realism \textit{pace} Carnap's linguistic realism.

\begin{description}
    \item[Ontology:] Are there wave functions? Within wave function theories, yes.
    \item[Epistemology:] Should we believe in its existence? Within wave function theories, yes.
    \item[Semantics:] Is the wave-function description true? Within wave function theories, yes.
\end{description}

There is strictly no room for belief in theoretical entities independently of the framework: to accept the framework is to accept its (internal) ontological commitments; if it is possible to say that there is, then, a belief, it is already built into the acceptance of the framework in question---thus blurring the line between the traditional debate over the realist's belief in the theory's truth \textit{versus} the (constructive) empiricist's acceptance of the theory's empirical adequacy \parencite[cf.][]{chakravarttyvanfraassen2018}.\footnote{\label{note:adequacy}One has to be cautious here when employing ``adequacy'', as this word is already a very loaded one in philosophy of science and can be interpreted in multiple ways. ``Adequacy'' as in ``latching'' the theory on the world and, in the other spectrum, as in ``empirical adequacy'' of van Fraassen's (\citeyear{vanfraassen1980}) constructive empiricism. This is a problem in itself which we'll leave for another occasion.} However, as science itself is in no position to say whether wave functions are \textit{indispensable} for doing quantum mechanics \parencite[cf.][]{wallace2021, bokulich2020, allori2020}, we should adopt an attitude of tolerance on these matters by suspending our judgement. Notice that this is the return of the problem of underdetermination, \textit{viz.} that we have frameworks for quantum theories that work with wave functions and frameworks for quantum theories that work with, say, point particles---and not wave functions. To our best knowledge, Carnap's account doesn't touch these matters. What matters to a Carnapian metaontology, from a methodological point of view, is its naturalistic guise, \textit{viz.} that ontology should be not only informed by science \parencite[cf.][]{maudlin2007, wallace2012}, but also without transgressing science \parencite[p.~87]{CARNAP1928atrans}---however ontological stuff it is! 

That is not to say that this is a simple task, science itself has a not total uncomplicated method to decide what is the \textit{best} theory, or how/why one adopts one among other options. There is a complex, and so far no totally explained way for the working scientist to make decisions on what theory to support What Carnap does \textit{not} endorse, however, is a full relativist sense of ontology. That is, as long as the ontological framework is tied up with science they are good to go, \textit{e.g.} whether with wave functions \parencite{ney2021} or point particles \parencite{bohmhiley2006}; with multiverses \parencite{wilson2020} or with causal consciousnesses \parencite{arroyoarenhart2019}. The same does not hold for an ontology of \textit{e.g.} unicorns! But we are not in a totally different position of unicorns as well, to maintain the example. Unicorns are not completely different from wave functions in the sense that linguistic realism prevents one from meaningfully asks whether \textit{e.g.} wave functions exist \textit{simpliciter}. Its naturalistic guise, however, prevents one from meaningfully ask whether, say, unicorns exist \textit{simpliciter}---as long as there are no current scientific theories on that kind of entity. In turn, we should stress that one can meaningfully ask whether, for example, \textit{phlogiston} exists within phlogiston theory in chemistry, and the answer would be ``yes''! The main difference between the phlogiston theory and electron theories for the linguistic realist is that the former is still in use.

With regard to the eventual product of language choice in relation to its suitability as a tool to reconstruct scientific theories, Carnap was not an advocate of arbitrariness. To see why, let us recall his ``Principle of Tolerance'' in \textcite{CARNAP1950}, which we divided into two parts:

\begin{quote}
    [1] Let us grant to those who work in any special field of investigation the freedom to use any form of expression which seems useful to them; [2] the work in the field will sooner or later lead to the elimination of those forms which have no useful function. \parencite[p.~40]{CARNAP1950}.
\end{quote}

Such a particular way of reading could be suggested by an interpretation limited to the first part of the Principle of Tolerance, isolated from the second---or if an exacerbated value was attributed only to the first operation at the expense of the second. On what concerns the second operation, the second part of the \textit{Principle} cannot be neglected for a satisfactory appraisal of Carnap's overall proposal, \textit{viz.} the task of assessing ``adequacy''\footnote{See note \ref{note:adequacy}.} to the linguistic alternative weighted in accordance with the established objectives.

When the second operation takes place, it significantly restricts the possible outcomes of the first. The second operation involves assessing the ``adequacy'' of these methods for understanding, in this case, the language of science, and this operation is not purely logical in nature, but also, globally, empirical.\footnote{Note that similar attitude appears in other parts of \textcite[p.~208]{CARNAP1950}, where the choice of the ``language of things'' is accounted for by its high degree of efficiency for most everyday statements. In this sense, as \textcite[42]{psillos2011} very well noted, there are agreements between Carnap and Feigl standpoints regarding these matters.} In other words, the freedom of the first operation is subsequently accounted for by the evaluation of the qualities of the forms of language with respect to the prescribed objectives. The exploration of these commitments and their consequences is functional for the demonstration of the best methods, the ``most appropriate'', for the explanation provided by the language of science. This, as Carnap insists, is a matter of pragmatical evaluations; and pragmatical evaluations are not detached from their intersubjective dependency. 

In this sense, the ``unlimited ocean of possible languages'' allowed by the Principle of Tolerance is not irresponsible/relativist. Despite its initial permissiveness in the construction of languages, these very language systems are always evaluated through their ability to model an adequate description of the object they are built for, \textit{viz.} science. Thus, Carnap writes in \textcite{CARNAP1939}:

\begin{quote}
    For any given calculus there are, in general, many different possibilities of a true interpretation. The practical situation, however, is such that for almost every calculus that is actually interpreted and applied in science, there is a certain interpretation or a certain kind of interpretation used in the great majority of cases of its practical application. This we will call the customary interpretation (or kind of interpretation) for the calculus. \parencite[p.~171]{CARNAP1939}.
\end{quote}

A Carnapian statement would be that the final choice between (say) two \textit{frameworks} is far from arbitrary, but even so, it is \textit{still} the product of a convention.\footnote{As Carnap pointed out in \textcite[fn.~4]{CARNAP1950}, this position is not dissimilar to that of \textcite[pp.~35--62]{FEIGL1950}.} This convention, however, is always accounted for according to the prescribed objectives. The freedom of choice provides the benefit that a completely deviant system may, in the future, prove useful to lay the groundwork for the language of science \parencite[cf.][p.~28]{CARNAP1939}, but this completely deviant system needs to be at the end of the day---and this is crucial---a \textit{functional} system.

What Carnap's linguistic realism cannot do is to specify a fact of the matter about what exists \textit{in reality}, so the realism is confined with existence questions that don't go beyond any given linguistic framework. In this sense, as the reader might already suspect, such a view is clearly very similar to Putnam's (\citeyear[chap.~3]{putnam1981}) so-called ``\textit{internal} realism''. And, as such, may fall prey to the same kind of criticism Putnam's internal realism did. For instance, here's \textcite{anderson1992}:

\begin{quote}
    Admittedly, Putnam's position does boast a rich ontology. Electrons exist every bit as much as chairs and tables do, and electrons can even help to \textit{explain} the superficial properties of macro-objects. Few realists, however, are willing to count this as a sufficient condition for being a ``realist.'' After all, Putnam insists that ontological commitment is always internal to a conceptual scheme; there is no scheme-independent fact of the matter about the ultimate furniture of the universe. \parencite[p.~49, original emphasis]{anderson1992}.
\end{quote}

Switch ``Putnam'' with ``Carnap'', and one would have the same kind of complaint going on within linguistic realism. So, to answer these kinds of questions is a pressing issue because, as we pointed out at the end of section \ref{sec:2}, the lack of a matter of fact on external questions is the reason why Carnap was framed as an anti-realist in the first place \parencite[cf.][]{chalmers2009}. But to be fair: who can do this? There don't seem to be, in fact, examples of scientific theories that deal with issues that concern \textit{the world} in fact, \textit{i.e.} regardless of the conceptual framework in which they operate---and admitting the opposite do not sound good, being somewhere in the spectrum between plainly wrong to epistemically unwarranted. Maybe bike workshops do that, but particle accelerators certainly don't. Let us press this point a little further. If the idea behind the external questions is related to the unity of science, then that is not really a problem. After all, \textit{e.g.} paediatrics and quantum mechanics have little/nothing to learn from each other, so one could say that they work with issues that are external to each other. But that is not what is at stake in realistic demand. What is at stake is reality. Thus, it seems to us, that the demand for external issues is not relatively external, but \textit{absolutely external} (i.e., external to all frameworks).

To exemplify such a claim, let us consider once again wave function realism. In the preface of \textcite{ney2021} (which consists of comprehensive development and defence of wave function \textit{realism}), \textcite{ney2021} acknowledges that there is no fact of the matter of which framework is the better one to understand quantum mechanics. This is why she calls the Carnapian notion of ``tolerance'':

\begin{quote}
\textelp{} while my main task here will be to make it clear that wave function realism is worth taking seriously as a framework for understanding the worlds described by our best quantum theories, \textit{my stance in this book will be one of humility and tolerance for other approaches}. \parencite[p.~\textsc{ix}, emphasis added]{ney2021}.
\end{quote}

So her defence of wave function realism is based on pragmatic criteria of this particular framework rather than on truth-conductive arguments. Furthermore, as it is well-known, there is no fact of the matter of which quantum theory is the right one, to begin with \parencite[cf.][]{durrlazarovici2020, arroyo-olegario2021}. Nevertheless, there are self-avowed ``realist'' approaches to quantum theories, such as the many-worlds realists with regard to Everettian quantum mechanics \parencite{wallace2012, wilson2020} and realists about Bohmian mechanics \parencite{bohmhiley2006}. Neither of such realists, however, are entitled to go beyond their internal questions of ontology and state something like ``\dots and that's how the \textit{world} is'' and no one seems to be calling off their \textit{realistic} attitude towards their own theory. How's that different from Carnapian linguistic realism? Put it bluntly under a conditional form: \textit{if} the above-mentioned self-avowed scientific realist approaches are realist indeed, \textit{then} the Carnapian approach can comfortably sit at the realist table as well. Here one might point out, as an anonymous referee did, that:

\begin{quote}
    The position of neutrality still seems to me to be the best option for interpreting the Carnapian theses on the subject in question here. However, Carnap's instrumentalist position in the partial interpretation of theoretical terms of axiomatic systems points to a position of denying semantic autonomy to such terms.
\end{quote}

However, for the reasons stated above, Carnap's position is not a completely neutral one.\footnote{Examples of complaint of such ``neutral'' reading of Carnap's metaontology can be found also in \textcite[p.~305]{Uebel2010}.}
That is, throughout this paper we saw that there is no such thing of semantic autonomy, so there cannot be this ``great realism'' in which the theoretical terms acquire their meaning \textit{because of nature}, \textit{viz.} independently of \textit{any} linguistic framework. However, Carnap endorses a form of realism \textit{for} and \textit{in} linguistic frameworks. As we mentioned, this became a familiar strategy employed by contemporary scientific realists, \textit{viz.} to refrain from external questions and endorse realism within specific linguistic/ontological frameworks. Whether or not this can be called ``scientific realism'' by hardcore realists is a question that we shall not strictly dwell on; rather, we proceed conditionally: \textit{if} Carnap's linguistic realism is indeed realist, then some contemporary, ``internal'' scientific realist approaches \parencite[cf.][]{psillos2011, ney2021} are realist as well.

In this sense, in order to entertain Carnap's linguistic realism, it seems that one needs to take models seriously and entertain, for example, what \textcite[p.~501]{schiemer2012} called ``truth in a model''---which can be another way to state truth \textit{within} a linguistic framework.

\section{Concluding remarks}\label{sec:5}

This article presented a conceptual clarification on the self-proclaimed position of Rudolf Carnap regarding the debate between realism and anti-realism in ontology and philosophy of science: the so-called ``linguistic realism''. Frequently, the Carnapian proposal is understood as deflationary about ontology, and anti-realist concerning scientific theories---in particular, to ontological commitments in relation to the theoretical terms of theories. 
We have moved away from this common interpretation articulating how Carnap's ``linguistic realism'' looks like, and how this discussion could elucidate recent proposals in the ontology of quantum mechanics such as the ``wave function realism''. As a result, we found that the Carnapian proposal has the motto of ``taking models seriously'', and the notable realist characteristics of adopting a framework---mainly, the ontological and semantic aspects---must be respectively understood as ``existing within a framework'' and ``true within a model''.\footnote{We must confess that we are refraining to use the expression ``frameworkism'' here only due to the ugliness of such expression!}

Properly understood, Carnap's positive approach to scientific realism is a \textit{linguistic}, \textit{ontological}, and \textit{internal} kind of realism. In this sense, it seems to resemble what \textcite[p.~65]{quine1951} called ``ontological commitment'', \textit{viz.} ``\textelp{} what, according to that theory, there is'', so this can be also a conciliatory way to look at the traditional Carnap--Quine debate. However, and this is another way of looking at it, linguistic realism could not have legitimate preferences for \textit{e.g.} desert landscapes, since this same (meta)ontological preference goes beyond the realm of science. Carnap's approach to ontology through the conception of linguistic frameworks is a self-avowed realist approach, namely, linguistic realism.

This, however, raises the question of whether this is realist \textit{enough}, which we think is the source of the numerous received views sketched at the end of section \ref{sec:2}. Some would say it is not,  \textit{pace} \textcite[chap.~3]{french2014} and \textit{contra} \textcite{psillos2012}. It would take (at least) a full paper to flesh out the sufficient and necessary conditions for a view to be called ``realist'', hence this discussion is (of course, alas!) beyond the scope of this one.


\section*{References}

{\small\frenchspacing

\noindent\hangindent5mm 
Allori, V. 2020. Scientific Realism without the Wave Function. In: S. French \& J. Saatsi (ed.), \textit{Scientific Realism and the Quantum}, p.212-228. Oxford: Oxford University Press.

\noindent\hangindent5mm 
Anderson, D. L. 1992. What Is Realistic about Putnam's Internal Realism?. \textit{Philosophical Topics}, 20(1): 49-83.

\noindent\hangindent5mm 
Arenhart, J. R. B. \& Arroyo, R. W. 2021a. Back to the question of ontology (and metaphysics). \textit{Manuscrito} 44(2): 1-51.

\noindent\hangindent5mm 
Arenhart, J. R. B. \& Arroyo, R. W. 2021b. On physics, metaphysics, and metametaphysics. \textit{Metaphilosophy} 52(2): 175-199.

\noindent\hangindent5mm 
Arroyo, R. W. \& Arenhart, J. R. B. 2019. Between physics and metaphysics: A discussion of the status of mind in quantum mechanics. In: J. A. de Barros \& C. Montemayor, \textit{Quanta and Mind: Essays on the Connection between Quantum Mechanics and the Consciousness}, p. 31-42. Cham: Springer.

\noindent\hangindent5mm 
Arroyo, R. W. \& da Silva, G. O. 2021. Against `Interpretation': Quantum Mechanics Beyond Syntax and Semantics. \textit{Axiomathes}: 1-37.

\noindent\hangindent5mm 
Bohm, D. \& Hiley, B. J. 2006. \textit{The Undivided Universe: An Ontological Interpretation of Quantum Theory}. London: Routledge.

\noindent\hangindent5mm 
Bokulich, A. 2020. Losing Sight of the Forest for the $\psi$: Beyond the Wavefunction Hegemony. In: S. French \& J. Saatsi (ed.), \textit{Scientific Realism and the Quantum}, p.185-211. Oxford: Oxford University Press.

\noindent\hangindent5mm 
Britannica. 2022. Deflate. \textit{Britannica English Dictionary}. https://www.britannica.com/dictionary/deflate. Access: 05.04.2022.

\noindent\hangindent5mm 
Bueno, O. 2016. Carnap, Logicism, and Ontological Commitment. In: S. Costreie (ed.), \textit{Early Analytic Philosophy: New Perspectives on the Tradition}, p. 337-352. Cham: Springer, p.

\noindent\hangindent5mm 
Cambridge University. 2022. Deflate, \textit{Cambridge English Dictionary}, https://dictionary.cambridge.org/dictionary/english/deflate. Access: 05.04.2022.

\noindent\hangindent5mm 
Carnap, R. 1931. Überwindung der Metaphysik durch logische Analyse der Sprache. \textit{Erkenntnis} 2(1): 219-241.

\noindent\hangindent5mm 
Carnap, R. 1938. Empiricism and the Language of Science, \textit{Synthese} 3(12): 33-35.

\noindent\hangindent5mm 
Carnap, R. 1939. Foundations of Logic and Mathematics. In: O. Neurath; R. Carnap; C. Morris (ed.), \textit{International Encyclopedia of Unified Science}, vol. 1, p. 139-213. Chicago:
University of Chicago Press.

\noindent\hangindent5mm 
Carnap, R. 1947. \textit{Meaning and Necessity. A Study in Semantics and Modal Logic}, 1st ed. Chicago: University of Chicago Press. Repr. \textit{Meaning and Necessity. A Study in Semantics and Modal Logic}, 2nd ed., Chicago: University of Chicago Press, 1956.

\noindent\hangindent5mm 
Carnap, R. 1950. Empiricism, Semantics and Ontology. \textit{Revue International de Philosophie}, 4. Reprint in: Carnap. R. 1956. \textit{Meaning and Necessity. A Study in Semantics and Modal Logic}, 2nd ed. Chicago: University of Chicago Press.

\noindent\hangindent5mm 
Carnap, R. 1963. Intellectual Autobiography. In: P. A. Schilpp (ed.), \textit{The Philosophy of Rudolf Carnap}, p.3-84. (The Library of Living Philosophers). Chicago: Open Court.

\noindent\hangindent5mm 
Carnap, R. 1966. \textit{Philosophical Foundations of Physics: An Introduction to the Philosophy of Science}. New York: Basic Books.

\noindent\hangindent5mm 
Carnap, R. 2003. Pseudoproblems in Philosophy: The Heteropsychological and the Realism Controversy. In: Carnap, R. \textit{The Logical Structure of the World [and] Pseudoproblems in Philosophy}, 2nd ed. Trans. by Rolf A. George. Chicago: Open Court.

\noindent\hangindent5mm 
Chakravartty, A. 2017. Scientific Realism. In: E. N. Zalta (ed.), \textit{The Stanford Encyclopedia of Philosophy}. Summer 2017 Edition. https://plato.stanford.edu/archives/sum2017/entries/scientific-realism/. Access: 05.04.2022.

\noindent\hangindent5mm 
Chakravartty, A. \& van Fraassen, B. C. 2018. What is Scientific Realism? \textit{Spontaneous Generations} 9(1): 12-25.

\noindent\hangindent5mm 
Chalmers, D. 2009. Ontological Anti-Realism. In: Chalmers, D.; Manley, D.; Wasserman, R. (ed.). \textit{Metametaphysics: New Essays on the Foundations of Ontology}, p.77-129. Oxford: Oxford University Press.

\noindent\hangindent5mm 
da Silva, G. O. 2020. \textit{Os empiristas vão à missa: compromissos ontológicos e frameworks linguísticos}. PhD thesis. Campinas: University of Campinas (UNICAMP). https://repositorio.unicamp.br/Acervo/Detalhe/1128755. Access: 05.04.2022.

\noindent\hangindent5mm 
Davidson, D. 1963. The method of extension and intension. In: P. A. Schilpp (ed.), \textit{The Philosophy of Rudolf Carnap}, p.311--349. (The Library of Living Philosophers). Chicago: Open Court, 1963.

\noindent\hangindent5mm 
Demopoulos, W. 2013. \textit{Logicism and its Philosophical Legacy}. Cambridge: Cambridge University Press.

\noindent\hangindent5mm 
Dürr, D. \& Lazarovici, D. 2020. \textit{Understanding Quantum Mechanics: The World According to Modern Quantum Foundations}. Cham: Springer, 2020

\noindent\hangindent5mm 
Einstein, A. 1921. \textit{Geometrie und Erfahrung: Erweiterte Fassung des Festvortrages Gehalten an der Preussischen Akademie der Wissenschaften zu Berlin am 27. Januar 1921}. Berlin: Springer.

\noindent\hangindent5mm 
Eklund, M. 2009. Carnap and Ontological Pluralism. In: Chalmers, D.; Manley, D.; Wasserman, R. (ed.). \textit{Metametaphysics: New Essays on the Foundations of Ontology}, p.130-156. Oxford:
Oxford University Press.

\noindent\hangindent5mm 
Falguera, J. L. \& Martínez-Vidal, C. 2020. Preface. In: Falguera, J. L. \& Martínez-Vidal, C. (ed.). \textit{Abstract Objects: For and Against}, p.v-xviii. Cham: Springer.

\noindent\hangindent5mm 
Feigl, H. 1950. Existential Hypotheses. Realistic versus Phenomenalistic Interpretations. \textit{Philosophy of Science} 17(1): 35-62.

\noindent\hangindent5mm 
French, S. 2014. \textit{The structure of the world: Metaphysics and representation}. Oxford: Oxford University Press.

\noindent\hangindent5mm 
Friedman, M. 2012. Carnap's Philosophical Neutrality Between Realism and Instrumentalism. In: Frappier, M.; Brown, D.; DiSalle, R. (ed.). \textit{Analysis and Interpretation in the Exact Sciences: Essays in honour of William Demopoulos}, p.95-114. Dordrecht: Springer.

\noindent\hangindent5mm 
Gentile, N. A. \& Gaeta, R. L. 2005. El neutralismo ontológico de Rudolf Carnap. \textit{V Jornadas de Investigación en Filosofía}: 1-10.

\noindent\hangindent5mm 
Hofweber, T. 2016. Carnap's Big Idea. In: S. Blatti \& S. Lapointe. (ed.). \textit{Ontology after Carnap}, p.13-30. Oxford: Oxford University Press.

\noindent\hangindent5mm 
Jaksland, R. 2020. Old problems for neo-positivist naturalized metaphysics. \textit{European Journal for Philosophy of Science} 10(16): 1-19.

\noindent\hangindent5mm 
Kraut, R. 2021. Rudolf Carnap: Pragmatist and expressivist about ontology. In: R. Bliss \& J. T. M. Miller (ed.). \textit{The Routledge Handbook of Metametaphysics}, p.32-48. London: Routledge.

\noindent\hangindent5mm 
Maudlin, T. 2019. \textit{Philosophy of physics: Quantum theory}. Princeton: Princeton University Press.

\noindent\hangindent5mm 
Maudlin, T. 2007. \textit{The metaphysics within physics}. Oxford: Oxford University Press.

\noindent\hangindent5mm 
Merriam-Webster. 2022. Deflate. \textit{Merriam-Webster.com dictionary}. https://www.merriam-webster.com/dictionary/deflate. Access: 05.04.2022.

\noindent\hangindent5mm 
Mizrahi, M. 2020. \textit{The Relativity of Theory: Key Positions and Arguments in the Contemporary Scientific Realism/Antirealism Debate}. Cham: Springer.

\noindent\hangindent5mm 
Neuber, M. 2014. Is Logical Empiricism Compatible With Scientific Realism? In: M. C. Galavotti; E. Nemeth; F. Stadler (ed.), \textit{European Philosophy of Science: Philosophy of Science in Europe and the Viennese Heritage}, p.249-262. Cham: Springer.

\noindent\hangindent5mm 
Ney, A. 2014. \textit{Metaphysics: an introduction}. New York: Routledge.

\noindent\hangindent5mm 
Ney, A. 2021. \textit{The World in the Wave Function}. Oxford: Oxford University Press.

\noindent\hangindent5mm 
Oxford University. 1989. \textit{The Oxford English Dictionary}, Vol. 4, CRE-DUZ. 2nd ed. Oxford: Oxford University Press. 

\noindent\hangindent5mm 
Parrini, P. 1994. With Carnap, beyond Carnap: Metaphysics, science, and the realism/instrumentalism controversy. In: W. Salmon and G. Wolters (ed.), \textit{Logic, Language, and the Structure of Scientific Theories}, p.255-277. Pittsburgh: Pittsburgh University Press.

\noindent\hangindent5mm 
Psillos, S. 2011. Choosing the realist framework. {Synthese} 180(2): 301-316.

\noindent\hangindent5mm 
Psillos, S. 2012. One Cannot Be Just a Little Bit Realist: Putnam and van Fraassen. In: J. R. Brown (ed.). \textit{Philosophy of Science: The Key Thinkers}, p;188-212. London: Continuum.

\noindent\hangindent5mm 
Psillos, S. 1999. \textit{Scientific realism: How science tracks truth}. London: Routledge.

\noindent\hangindent5mm 
Putnam, H. 1981. \textit{Reason, Truth and History}. Cambridge: Cambridge University Press.

\noindent\hangindent5mm 
Putnam, H. 1975. X*--What is ``Realism''? \textit{Proceedings of the Aristotelian Society} 76(1): 177-194.

\noindent\hangindent5mm 
Quine, W. v. O. 1951. On Carnap's Views on Ontology. \textit{Philosophical Studies} 2(5): 65-72.

\noindent\hangindent5mm 
Schiemer, G. 2012. Carnap's early semantics. \textit{Erkenntnis} 78(3): 487-522.

\noindent\hangindent5mm 
Thomasson, A. 2016. Carnap and the Prospects for Easy Ontology. In: S. Blatti \& S. Lapointe, S. (ed.). \textit{Ontology after Carnap}, p.122-144. Oxford: Oxford University Press.

\noindent\hangindent5mm 
Torfehnezhad, P. 2017. In Carnap's Defense: A survey on the concept of a linguistic framework in Carnap's philosophy. \textit{Abstracta} 9(1): 3-30.

\noindent\hangindent5mm 
Uebel, T. 2010. Carnap and the Perils of Ramseyfication. In: M. Suárez; M. Dorato; M. Rédei (ed.), \textit{EPSA Epistemology and Methodology of Science: Launch of the European Philosophy of Science Association}, p.299-310. Dordrecht: Springer.

\noindent\hangindent5mm 
van Fraassen, B. C. 1980. \textit{The Scientific Image}. Oxford: Oxford University Press.

\noindent\hangindent5mm 
Wallace, D. 2021. Against Wavefunction Realism. In: S. Dasgupta; R. Dotan; B. Weslake (ed.). \textit{Current Controversies in Philosophy of Science}, p.63-74. New York: Routledge.

\noindent\hangindent5mm 
Wallace, D. 2012. \textit{The emergent multiverse: Quantum theory according to the Everett interpretation}. Oxford: Oxford University Press.

\noindent\hangindent5mm 
Wilson, A. 2020. \textit{The Nature of Contingency: Quantum Physics as Modal Realism}. Oxford: Oxford University Press.

}


\selectlanguage{english}

\begingroup
\parindent 0pt
\def\enotesize{\footnotesize}
\theendnotes
\endgroup


\subsection*{Acknowledgments}


{\small

Order of authorship is alphabetical and does not represent any kind of priority; authors contributed equally to this work. A preliminary version of this article was presented by Gilson Olegario da Silva at the \textit{12th Principia International Symposium}, 2021, under the title ``Realism without metaphysics'' (in Portuguese).

}


@article{FEIGL1950,
	author = {Feigl, Herbert},
	doi = {10.1086/287065},
	issn = {0031-8248, 1539-767X},
	journal = {Philosophy of Science},
	number = {1},
	pages = {35-62},
	publisher = {University of Chicago Press},
	source = {Crossref},
	title = {Existential Hypotheses. Realistic versus Phenomenalistic Interpretations},
	volume = {17},
	year = {1950}}

@book{demopoulos2013,
	Author = {Demopoulos, William},
	Publisher = {Cambridge University Press},
	Title = {Logicism and its Philosophical Legacy},
	Year = {2013}
}

@book{psillos1999,
	Address = {London \& New York},
	Author = {Psillos, Stathis},
	Publisher = {Routledge},
	Title = {Scientific realism: How science tracks truth},
	Year = {1999}
}

@book{vanfraassen1980,
	Address={Oxford},
	Author = {van{\ }Fraassen, Bas C.},
	Publisher = {Oxford University Press},
	Title = {The Scientific Image},
	Year = {1980}
}

@incollection{friedman2012a,
	Author = {Friedman, Michael},
	Booktitle = {Analysis and Interpretation in the Exact Sciences: Essays in honour of William Demopoulos},
	Editor = {Frappier, Melanie and Brown, Derek and Di{}Salle, Robert},
	Pages = {95--114},
	Publisher = {Springer Netherlands},
	Series = {The Western Ontario Series in Philosophy of Science},
	Title = {{Carnap's} Philosophical Neutrality Between Realism and Instrumentalism},
	Volume = {78},
	Year = {2012}
}

@inproceedings{gentile2005,
	Author = {Gentile, N{\'e}lida Alcira and Gaeta, Rodolfo Luj{\'a}n},
	Booktitle = {V Jornadas de Investigaci{\'o}n en Filosof\'ia},
	Title = {El neutralismo ontol{\'o}gico de Rudolf Carnap},
	Year = {2005}
}

@incollection{bueno2016,
author="Otávio Bueno",
editor="Sorin Costreie",
title="Carnap, Logicism, and Ontological Commitment",
bookTitle="Early Analytic Philosophy: New Perspectives on the Tradition",
year="2016",
publisher="Springer",
address="Cham",
pages="337--352"
}

@incollection{chalmers2009,
author="David Chalmers",
editor="David Chalmers and David Manley and Ryan Wasserman",
title="Ontological Anti-Realism",
bookTitle="Metametaphysics: New Essays on the Foundations of Ontology",
year="2009",
publisher="Oxford University Press",
address="Oxford",
pages="77--129"
}

@incollection{eklund2009,
author="Matti Eklund",
editor="David Chalmers and David Manley and Ryan Wasserman",
title="Carnap and Ontological Pluralism",
bookTitle="Metametaphysics: New Essays on the Foundations of Ontology",
year="2009",
publisher="Oxford University Press",
address="Oxford",
pages="130--156"
}

@InCollection{chakravartty2017sep,
	author       =	{Anjan Chakravartty},
	title        =	{{Scientific Realism}},
	booktitle    =	{The {Stanford} Encyclopedia of Philosophy},
	editor       =	{Edward N. Zalta},
	howpublished =	{\url{https://plato.stanford.edu/archives/sum2017/entries/scientific-realism/}},
	year         =	{2017},
	edition      =	{{S}ummer 2017},
	publisher    =	{Metaphysics Research Lab, Stanford University}
}

@article{putnam1975,
author={Hilary Putnam},
title={X*---{W}hat is ``{R}ealism''?},
journal={Proceedings of the Aristotelian Society},
volume={76},
number = {1},
pages={177--194},
year={1975}
}

@book{mizrahi2020,
author={Moti Mizrahi},
title={The Relativity of Theory: Key Positions and Arguments in the Contemporary Scientific Realism/Antirealism Debate},
series={Synthese Library},
volume={431},
address={Cham},
publisher={Springer},
year={2020}
}

@article{arroyo-olegario2021,
title={{Against `Interpretation': Quantum Mechanics Beyond Syntax and Semantics}},
author={Raoni Wohnrath Arroyo and da{\ }Silva, Gilson Olegario},
journal={Axiomathes},
pages={1--37},
doi={10.1007/s10516-021-09579-y},
year={2021}
}

@book{ney2021,
author={Alyssa Ney},
title={The World in the Wave Function},
address={Oxford},
publisher={Oxford University Press},
year={2021}
}

@book{maudlin2019,
    author    = "Maudlin, Tim",
    title     = "Philosophy of physics: Quantum theory",
    publisher = "Princeton University Press",
    series   = "Princeton Foundations of Contemporary Philosophy",
    address  = "Princeton",
    year      = "2019"
}

@incollection{bokulich2020,
booktitle={Scientific Realism and the Quantum},
editor={Steven French and Juha Saatsi},
publisher={Oxford University Press},
address={Oxford},
year={2020},
author={Alisa Bokulich},
title={{Losing Sight of the Forest for the $\psi$: Beyond the Wavefunction Hegemony}},
pages={185--211}
}

@incollection{allori2020,
booktitle={Scientific Realism and the Quantum},
editor={Steven French and Juha Saatsi},
publisher={Oxford University Press},
address={Oxford},
year={2020},
author={Valia Allori},
title={{Scientific Realism without the Wave Function}},
pages={212--228}
}

@article{arenhartarroyo2021meta,
author={Jonas R. Becker Arenhart and Raoni Wohnrath Arroyo},
title={On physics, metaphysics, and metametaphysics},
journal={Metaphilosophy},
year={2021},
volume={52},
number={2},
pages={175--199}
}

@book{putnam1981,
author={Hilary Putnam},
title={Reason, Truth and History},
address={Cambridge},
publisher={Cambridge University Press},
year={1981}
}

@incollection{thomasson2016,
	Address = {Oxford},
	Author = {Amie Thomasson},
	Booktitle = {Ontology after {C}arnap},
	Editor = {Stephan Blatti and Sandra Lapointe},
	Pages = {122--144},
	Publisher = {Oxford University Press},
	Title = {Carnap and the Prospects for Easy Ontology},
	Year = {2016}
}

@book{ney2014,
title={Metaphysics: an introduction},
author={Alyssa Ney},
publisher={Routledge},
address={New York},
year={2014}
}

@incollection{hofweber2016,
	Address = {Oxford},
	Author = {Thomas Hofweber},
	Booktitle = {Ontology after {C}arnap},
	Editor = {Stephan Blatti and Sandra Lapointe},
	Pages = {13--30},
	Publisher = {Oxford University Press},
	Title = {Carnap's Big Idea},
	Year = {2016}
}

@article{arenhartarroyo2021manu,
Author={Jonas R. Becker Arenhart and Raoni Wohnrath Arroyo},
title={Back to the question of ontology (and metaphysics)},
journal={Manuscrito},
volume={44},
number={2},
pages={1--51},
year={2021}
}

@incollection{falguera-martinezvidal2020pref,
author={Jos\'e L. Falguera and Concha Mart\'inez{-}Vidal},
editor={Jos\'e L. Falguera and Concha Mart\'inez{-}Vidal},
booktitle={Abstract Objects: For and Against},
title={Preface},
pages={v--xviii},
publisher={Springer},
address={Cham},
year={2020}
}

@inbook{CARNAP1928btrans,
	author = {Carnap, Rudolf},
	booktitle = {The Logical Structure of the World [and] Pseudoproblems in Philosophy},
	crossref = {CARNAP1928atrans},
	title = {Pseudoproblems in Philosophy: The Heteropsychological and the Realism Controversy},
	translator = {Rolf A. George},
	year = {2003}
}

@book{CARNAP1928atrans,
	author = {Carnap, Rudolf},
	edition = {2},
	location = {Chicago},
	publisher = {Open Court},
	title = {The Logical Structure of the World [and] Pseudoproblems in Philosophy},
	translator = {Rolf A. George},
	year = {2003}
}

@article{CARNAP1950,
	author = {Carnap, Rudolf},
	journal = {Revue International de Philosophie},
	title = {Empiricism, Semantics and Ontology},
	volume = {4},
	year = {1950}
}

@incollection{kraut2021rudolf,
  title={Rudolf Carnap: Pragmatist and expressivist about ontology},
  author={Kraut, Robert},
  booktitle={The Routledge Handbook of Metametaphysics},
  editor={Ricki Bliss and J. T. M. Miller},
  pages={32--48},
  address={London},
  year={2021},
  publisher={Routledge}
}

@Incollection{Neuber2014,
author="Neuber, Matthias",
editor="Galavotti, Maria Carla and Nemeth, Elisabeth and Stadler, Friedrich",
title="Is Logical Empiricism Compatible With Scientific Realism?",
bookTitle="European Philosophy of Science -- Philosophy of Science in Europe and the Viennese Heritage",
year="2014",
publisher="Springer International Publishing",
address="Cham",
pages="249--262"
}

@article{torfehnezhad2017carnap,
	author = {Torfehnezhad, Parzhad},
	journal = {Abstracta},
	number = {1},
	title = {In Carnap's Defense: A survey on the concept of a linguistic framework in Carnap's philosophy},
	volume = {9},
	year = {2017}}

@incollection{CARNAP1939,
	author = {Carnap, Rudolf},
	booktitle = {International Encyclopedia of Unified Science},
	editor = {Neurath, Otto and Carnap, Rudolf and Morris, Charles},
	location = {Chicago},
	number = {3},
	pages = {139--213},
	publisher = {University of Chicago Press},
	title = {Foundations of Logic and Mathematics},
	volume = {1},
	year = {1939}
}

@incollection{CARNAP1963a,
	author = {Carnap, Rudolf},
	booktitle = {The Philosophy of {R}udolf {C}arnap},
	crossref = {CARNAP1963-orig},
	pages = {3--84},
	title = {Intellectual Autobiography},
	year = {1963}}

@inbook{Einstein1921,
	address = {Berlin, Heidelberg},
	author = {Einstein, Albert},
	booktitle = {Geometrie und Erfahrung: Erweiterte Fassung des Festvortrages Gehalten an der Preussischen Akademie der Wissenschaften zu Berlin am 27. Januar 1921},
	chapter = {Geometrie und Erfahrung},
	pages = {2--20},
	publisher = {Springer Berlin Heidelberg},
	source = {Crossref},
	title = {Geometrie und Erfahrung},
	year = {1921}
	}

@book{CARNAP1966,
	author = {Carnap, Rudolf},
	location = {New York},
	note = {Editor Martin Gardner},
	publisher = {Basic Books},
	title = {Philosophical Foundations of Physics: An Introduction to the Philosophy of Science},
	year = {1966}
}

@phdthesis{olegariotese2020,
	address = {Campinas},
	author = {da{\ }Silva, Gilson Olegario},
	school = {University of Campinas (UNICAMP)},
	title = {Os empiristas v{\~a}o \`{a} missa: compromissos ontol\'{o}gicos e frameworks lingu\'{i}sticos},
	url = {http://repositorio.unicamp.br/jspui/handle/REPOSIP/342469},
	year = {2020}
	}

@article{CARNAP1938b,
	author = {Rudolf Carnap},
	journal = {Synthese},
	number = {12},
	pages = {33--35},
	title = {Empiricism and the Language of Science},
	volume = {3},
	year = {1938}
	}

@article{quine1951,
 author = {Quine, Willard van Omar},
 journal = {Philosophical Studies},
 number = {5},
 pages = {65--72},
 title = {On {C}arnap's Views on Ontology},
 volume = {2},
 year = {1951}
}

@incollection{wallace2021,
	Address = {New York},
	Author = {David Wallace},
	Booktitle = {Current Controversies in Philosophy of Science},
	Editor = {Shamik Dasgupta and Ravit Dotan and Brad Weslake},
	Pages = {63--74},
	Publisher = {Routledge},
	Title = {{Against Wavefunction Realism}},
	Year = {2021}
}

@book{maudlin2007,
	Address = {Oxford},
	Author = {Maudlin, Tim},
	Publisher = {Oxford University Press},
	Title = {The metaphysics within physics},
	Year = {2007}}

@book{wallace2012,
	Address = {Oxford},
	Author = {Wallace, David},
	Publisher = {Oxford University Press},
	Title = {{The emergent multiverse: Quantum theory according to the Everett interpretation}},
	Year = {2012}}

@book{bohmhiley2006,
	Address = {London},
	Author = {Bohm, David and Hiley, Basil J.},
	Publisher = {Routledge},
	Title = {{The Undivided Universe: An Ontological Interpretation of Quantum Theory}},
	Year = {2006}}

@book{wilson2020,
	Address = {Oxford},
	Author = {Wilson, Alastair},
	Publisher = {Oxford University Press},
	Title = {The Nature of Contingency: Quantum Physics as Modal Realism},
	Year = {2020}}

@incollection{arroyoarenhart2019,
	Address = {Cham},
	Author = {Arroyo, Raoni Wohnrath and Jonas R. Becker Arenhart},
	Booktitle = {{Quanta and Mind: Essays on the Connection between Quantum Mechanics and the Consciousness}},
	Editor = {de{\ }Barros, Jos\'e Acacio and Montemayor, Carlos},
	Pages = {31--42},
	Publisher = {Springer},
	Title = {{Between physics and metaphysics: A discussion of the status of mind in quantum mechanics}},
	Year = {2019}}

@book{durrlazarovici2020,
author={Detlef D\"urr and Dustin Lazarovici},
title={Understanding Quantum Mechanics: The World According to Modern Quantum Foundations},
publisher={Springer},
address={Cham},
year={2020}
}

@article{anderson1992,
title={{What Is Realistic about Putnam's Internal Realism?}},
author={David L. Anderson},
journal={Philosophical Topics},
volume={20},
number={1},
pages={49--83},
year={1992}
}

@article{schiemer2012,
title={{Carnap's early semantics}},
author={Georg Schiemer},
journal={Erkenntnis},
volume={78},
number={3},
pages={487--522},
year={2012}
}

@book{french2014,
	Address = {Oxford},
	Author = {Steven French},
	Publisher = {Oxford University Press},
	Title = {{The structure of the world: Metaphysics and representation}},
	Year = {2014}
}

@book{CARNAP1947,
	author = {Carnap, Rudolf},
	edition = {1},
	location = {Chicago},
	publisher = {University of Chicago Press},
	subtitle = {A Study in Semantics and Modal Logic},
	title = {Meaning and Necessity},
	year = {1947}
}

@article{psillos2011,
	author = {Psillos, Stathis},
	journal = {Synthese},
	number = {2},
	pages = {301--316},
	title = {Choosing the realist framework},
	volume = {180},
	year = {2011}
	}

@article{CARNAP1931,
	author = {Carnap, Rudolf},
	journal = {Erkenntnis},
	number = {1},
	pages = {219--241},
	publisher = {Springer Netherlands},
	title = {{U}berwindung der {M}etaphysik durch logische {A}nalyse der {S}prache},
	volume = {2},
	year = {1931}
}

@incollection{CARNAP1963-orig,
	author = {Carnap, Rudolf},
	editor = {Schilpp, Paul Arthur},
	location = {Chicago},
	publisher = {Open Court},
	series = {The Library of Living Philosophers},
	title = {{T}he Philosophy of {R}udolf {C}arnap},
	year = {1963}
}

@inbook{davidson1963method,
	author = {Davidson, Donald},
	crossref = {CARNAP1963-orig},
	title = {The method of extension and intension},
	year = {1963}}

@incollection{parrini1994carnap,
	author = {Parrini, Paolo},
	booktitle = {Logic, Language, and the Structure of Scientific Theories},
	editor = {Wesley C. Salmon, Gereon Wolters},
	journal = {Logic, Language, and the Structure of Scientific Theories, Universitatsverlag Konstanz und Pittsburgh University Press, Pittsburgh und Konstanz},
	location = {Pittsburgh Und Konstanz},
	pages = {255--277},
	publisher = {Universitätsverlag Konstanz Und Pittsburgh University Press},
	title = {With Carnap, beyond Carnap: Metaphysics, science, and the realism/instrumentalism controversy},
	year = {1994}}

@incollection{psillos2012,
author={Stathis Psillos},
title={{One Cannot Be Just a Little Bit Realist: Putnam and van Fraassen}},
pages={188--212},
booktitle={{Philosophy of Science: The Key Thinkers}},
editor={James Robert Brown},
address={London},
publisher={Continuum},
year={2012}
}

@article{chakravarttyvanfraassen2018,
title={What is Scientific Realism?},
author={Chakravartty, Anjan and van{\ }Fraassen, Bas C.},
journal={Spontaneous Generations},
volume={9},
number={1},
pages={12--25},
year={2018}
}

@article{jaksland2020,
author={Rasmus Jaksland},
title={Old problems for neo-positivist naturalized metaphysics},
journal={European Journal for Philosophy of Science},
volume={10},
number={16},
pages={1--19},
year={2020}
}

@misc{mw:deflate,
  author    = {Merriam{-}Webster},
  howpublished = {Merriam-Webster.com dictionary},
  url       = {https://www.merriam-webster.com/dictionary/deflate},
  title     = {Deflate},
  year   = {2022},
}

@misc{ced:deflate,
  author    = {Cambridge{\ }University},
  title     = {Deflate},
  howpublished = {Cambridge English Dictionary},
  url       = {https://dictionary.cambridge.org/dictionary/english/deflate},
  year   = {2022},
}

@misc{bd:deflate,
  author    = {Britannica},
  howpublished = {Britannica English Dictionary},
  url       = {https://www.britannica.com/dictionary/deflate},
  title     = {Deflate},
  year   = {2022},
}

@book{oed1989,
author={Oxford{\ }University},
title={The Oxford English Dictionary},
edition={2},
volume={4, CRE-DUZ},
publisher={Oxford University Press},
address={Oxford},
year={1989}
}

@inbook{Uebel2010,
	address = {Dordrecht},
	author = {Uebel, Thomas},
	booktitle = {EPSA Epistemology and Methodology of Science: Launch of the European Philosophy of Science Association},
	editor = {Suárez, Mauricio and Dorato, Mauro and R\'edei, Mikl\'os},
	isbn = {978-90-481-3263-8},
	pages = {299--310},
	publisher = {Springer Netherlands},
	title = {Carnap and the Perils of Ramseyfication},
	year = {2010}}
\end{document}